# Alternative formalism to the slave particle mean field theory of the t-J model without deconfinement

**Takao Morinari**

Yukawa Institute for Theoretical Physics, Kyoto University Kyoto 606-8502, Japan

E-mail: `morinari@yukawa.kyoto-u.ac.jp`

**Abstract.** An alternative formalism that does not require the assumption of the deconfinement phase of a U(1) gauge field is proposed for the slave particle mean field theory. Starting form the spin-fermion model, a spinon field, which is either fermion or boson, is introduced to represent the localized spin moment. We find a d-wave superconductive state in the mean field theory in the case of the fermion representation of the localized spin moment that corresponds to the slave boson mean field theory of the t-J model, whereas the d-wave superconductive state is absent in case of the Schwinger boson representation of the localized spin moments.

## 1. Introduction

As a theory for high-temperature superconductivity, the slave particle mean field theory of the t-J model has been studied intensively [1, 2, 3, 4, 5]. In the slave particle formalism, the electron operator is expressed in terms of auxiliary fermions and bosons. For instance, in the slave boson formalism the electron annihilation operator $c_{j\sigma}$ at site $j$ with spin $\sigma$ is given by

$$c_{j\sigma} = b_j^\dagger f_{j\sigma}. \qquad (1)$$

Here $b_j^\dagger$ is a boson operator and $f_{j\sigma}$ is a fermion operator. The boson created by $b_j^\dagger$ is called holon and is supposed to carry the electron's charge. While, the fermion created by $f_{j\sigma}^\dagger$ is called spinon and is supposed to carry the electron's spin and carry no charge.

Application of the slave boson formalism to the t-J model suggests a simple and attractive picture for $d_{x^2-y^2}$-wave superconductivity within a mean field theory. The $d_{x^2-y^2}$-wave superconductive state is described as the state with the spinon pairing with $d_{x^2-y^2}$-wave symmetry and the holon Bose-Einstein condensation[3, 4]. The state with the same spinon pairing without holon condensation is suggested to be a pseudogap phase[6]. U(1) gauge field fluctuations around the mean field state lead to non-Fermi liquid like behaviors, such as $T$-linear resistivity law[7, 8, 9].

Although the picture suggested by the slave boson mean field theory of the t-J model is intriguing, the theory is based on a crucial assumption. This assumption is related to the invariance of the electron operators under the U(1) gauge transformation of $b_j \to b_j \exp(i\theta_j)$ and $f_{j\sigma} \to f_{j\sigma} \exp(i\theta_j)$. The assumption of the theory is that the U(1) gauge field associated with this gauge symmetry is in the deconfined phase. If the U(1) gauge field is in the deconfined phase, then the holons and the spinons are independent particles. In other words, the system is in the spin-charge separation phase[10]. However, Polyakov showed that the pure compact U(1) gauge field theory is always confining[11]. The situation is quite unclear when there is matter fields as in the slave particle theory. Nagaosa argued that dissipation effects associated with the presence of the Fermi surface of the spinons can lead to the deconfined phase through suppression of instanton proliferation[12]. However, in Ref.[13] it is argued that this mechanism can be suppressed by screening effect. Another approach based on duality mapping suggests that the system is confining[14]. The issue on the presence of the deconfined phase of the U(1) gauge field theory is still in controversial [15, 16, 17, 18].

In this paper, we will not try directly to examine the possibility of the deconfined phase within the slave particle mean field theory of the t-J model. Instead, we derive a similar theory by taking a different formalism. We start with the spin-fermion model[19, 20, 21, 22]. If one takes the strong coupling limit, then the model is reduced to the t-J model. In the slave particle formalism, fermion and boson fields are introduced at the t-J model. Here we introduce them before taking the strong coupling limit. Since the carrier fields and the localized spin moments are independent degrees of freedom, we can express the localized spin moment either by a fermion operator or by a boson operator as independent fields. So it is not necessary to assume the deconfined phase to make the fields independent degrees of freedom.

## 2. Alternative formalism for the slave boson mean field theory

The model may be given by

$$H = -t\sum_{\langle i,j \rangle}\left(c_i^\dagger c_j + c_j^\dagger c_i\right) + \frac{1}{2}J_K\sum_j\left(c_j^\dagger \boldsymbol{\sigma} c_j\right)\cdot \mathbf{S}_j + J\sum_{\langle i,j \rangle}\mathbf{S}_i\cdot\mathbf{S}_j. \quad (2)$$

Here the carrier field is represented by a spinor $c_i^\dagger = \left(c_{i\uparrow}^\dagger, c_{i\downarrow}^\dagger\right)$. The components of the vector $\boldsymbol{\sigma} = (\sigma_1, \sigma_2, \sigma_3)$ are the Pauli spin matrices. The field $\mathbf{S}_j$ describes the localized spin at site $j$. Here, for simplicity, we do not distinguish the oxygen lattice sites and the copper lattice sites. (More precisely, $c_{i\sigma}^\dagger$ create states represented by the Wannier function of the symmetric state of four oxygen sites around a copper ion[23].)

In the model, the largest parameter is $J_K$, which leads to a picture of the Zhang-Rice singlet[23]. If one takes the $J_K \to \infty$ limit, then the model is reduced to the t-J model. In the slave particle formalism, fermion and boson fields are introduced after taking the $J_K \to \infty$ limit. Here we first introduce either a fermion or a boson field that is independent of the carrier field in order to describe the localized spin moment. The $J_K \to \infty$ limit shall be considered at the final step.

There are several ways to represent the localized spin degrees of freedom. Here we consider the Abrikosov's pseudofermion representation and the Schwinger boson representation. In the former representation, the localized spin $\mathbf{S}_j$ is described by

$$\mathbf{S}_j = \frac{1}{2}f_j^\dagger \boldsymbol{\sigma} f_j, \quad (3)$$

with the constraint $\sum_\sigma f_{j\sigma}^\dagger f_{j\sigma} = 1$. In the path-integral formalism, the partition function is given by $\mathcal{Z} = \int \mathcal{D}\bar{c}\mathcal{D}c\mathcal{D}\bar{f}\mathcal{D}f\mathcal{D}\lambda \exp(-S)$, where

$$\begin{aligned}S = \int_0^\beta d\tau &\left[\sum_j \bar{c}_j(\partial_\tau - \mu)c_j + \sum_j \bar{f}_j \partial_\tau f_j \right.\\
&+ \sum_j \lambda_j\left(\sum_\sigma \bar{f}_{j\sigma}f_{j\sigma} - 2\tilde{S}\right) - t\sum_{\langle i,j\rangle}(\bar{c}_i c_j + \bar{c}_j c_i)\\
&\left. + \frac{J_K}{4}\sum_j (\bar{c}_j \boldsymbol{\sigma} c_j)\cdot(\bar{f}_j \boldsymbol{\sigma} f_j) + \frac{J}{4}\sum_{\langle i,j\rangle}(\bar{f}_i \boldsymbol{\sigma} f_i)\cdot(\bar{f}_j \boldsymbol{\sigma} f_j)\right].\end{aligned} \quad (4)$$

Due to strong Kondo coupling $J_K$, the carriers can combine with the localized spin moments to form a singlet pair[23]. This picture is justified in the $J_K \to \infty$ limit. In order to take into account this correlation effect, we rewrite the Kondo coupling term using Stratonovich-Hubbard (SH) fields as follows: $(\bar{c}_j \boldsymbol{\sigma} c_j)\cdot(\bar{f}_j \boldsymbol{\sigma} f_j) \to \frac{3}{2}\bar{s}_j s_j - \frac{3}{\sqrt{2}}\bar{s}_j(f_{j\downarrow}c_{j\uparrow} - f_{j\uparrow}c_{j\downarrow}) - \frac{3}{\sqrt{2}}s_j(\bar{c}_{j\uparrow}\bar{f}_{j\downarrow} - \bar{c}_{j\downarrow}\bar{f}_{j\uparrow})$. Here we have introduced the SH fields $s_j$ and $\bar{s}_j$ that describe the singlet pair formation between the carrier field and the pseudofermion. Note that the SH field for the triplet pair formation has been omitted because the mean field values of them vanish at the saddle point reflecting that the Kondo coupling is antiferromagnetic.

Our next task is to rewrite the Heisenberg term. For the Heisenberg term in Eq.(2), we introduce the SH fields $D_{ij}$ and $\overline{D}_{ij}$ as follows, $(\overline{f}_i \boldsymbol{\sigma} f_i) \cdot (\overline{f}_j \boldsymbol{\sigma} f_j) \to 2\overline{D}_{ij} D_{ij} - 2\overline{D}_{ij}(f_{i\uparrow} f_{j\downarrow} - f_{i\downarrow} f_{j\uparrow}) - 2 D_{ij}(\overline{f}_{j\downarrow}\overline{f}_{i\uparrow} - \overline{f}_{j\uparrow}\overline{f}_{i\downarrow})$. The field $D_{ij}$ is identical to the mean field taken for d-wave pairing of spinons in the slave boson mean field theory of the t-J model[3, 4]. In addition, we introduce the following SH fields $\overline{\chi}_{ij}$ and $\chi_{ij}$: $(\overline{f}_i \boldsymbol{\sigma} f_i) \cdot (\overline{f}_j \boldsymbol{\sigma} f_j) \to 2\overline{\chi}_{ij}\chi_{ij} - 2\overline{\chi}_{ij}(\overline{f}_{i\uparrow} f_{j\uparrow} + \overline{f}_{i\downarrow} f_{j\downarrow}) - 2\chi_{ij}(\overline{f}_{j\uparrow} f_{i\uparrow} + \overline{f}_{j\downarrow} f_{i\downarrow})$. The field $\chi_{ij}$ and $\overline{\chi}_{ij}$ are identical to the mean fields taken by Nagaosa and Lee[7, 8] to describe the normal state properties of the system.

We consider the states with uniform values of $s_j = s_0$, $D_{i,i+\hat{x}} = -D_{i,i+\hat{y}} = D_0$, and $\chi_{ij} = \chi_0$. In the momentum and the Matsubara frequency space, the action is given by

$$S = \sum_k \left[ \begin{pmatrix} \overline{c}_{k\uparrow} & c_{-k\downarrow} \end{pmatrix} \begin{pmatrix} -i\omega_n + \xi_\mathbf{k} & 0 \\ 0 & -i\omega_n - \xi_\mathbf{k} \end{pmatrix} \begin{pmatrix} c_{k\uparrow} \\ \overline{c}_{-k\downarrow} \end{pmatrix} \right.$$
$$+ \begin{pmatrix} \overline{f}_{k\uparrow} & f_{-k\downarrow} \end{pmatrix} \begin{pmatrix} -i\omega_n + \chi_\mathbf{k} + \lambda & \Delta_\mathbf{k} \\ \overline{\Delta}_\mathbf{k} & -i\omega_n - \chi_\mathbf{k} - \lambda \end{pmatrix} \begin{pmatrix} f_{k\uparrow} \\ \overline{f}_{-k\downarrow} \end{pmatrix}$$
$$+ \begin{pmatrix} \overline{c}_{k\uparrow} & c_{-k\downarrow} \end{pmatrix} \begin{pmatrix} 0 & -\eta \\ \overline{\eta} & 0 \end{pmatrix} \begin{pmatrix} f_{k\uparrow} \\ \overline{f}_{-k\downarrow} \end{pmatrix}$$
$$\left. + \begin{pmatrix} \overline{f}_{k\uparrow} & f_{-k\downarrow} \end{pmatrix} \begin{pmatrix} 0 & \eta \\ -\overline{\eta} & 0 \end{pmatrix} \begin{pmatrix} c_{k\uparrow} \\ \overline{c}_{-k\downarrow} \end{pmatrix} \right] \quad (5)$$

with $k = (\mathbf{k}, i\omega_n)$, $\xi_\mathbf{k} = -2t(\cos k_x + \cos k_y) - \mu$, $\chi_\mathbf{k} = -\chi_0 J(\cos k_x + \cos k_y)$, $\Delta_\mathbf{k} = JD_0(\cos k_x - \cos k_y)$, and $\eta = 3\sqrt{2}J_K s_0/4$. The characteristic temperature for $D_0$ is on the order of $J$, which is identical to that of the slave boson mean field theory of the t-J model. For $s_0$, the characteristic temperature is given by $k_B T \sim W \exp(-1/(J_K N_F))$ with $W$ the band width and $N_F$ the density of states at the Fermi surface of the conduction electrons within the weak coupling theory. While, $k_B T \sim J_K$ in the strong coupling limit. By integrating out the fermion fields $\overline{f}_j$ and $f_j$, we find

$$S = \sum_k \begin{pmatrix} \overline{c}_{k\uparrow} & c_{-k\downarrow} \end{pmatrix} \left[ \begin{pmatrix} -i\omega_n + \xi_k & 0 \\ 0 & -i\omega_n - \xi_k \end{pmatrix} \right.$$
$$+ \frac{1}{(i\omega_n)^2 - (\chi_k + \lambda)^2 - |\Delta_k|^2}$$
$$\left. \times \begin{pmatrix} \overline{\eta}\eta(-i\omega_n + \chi_k + \lambda) & \eta^2 \overline{\Delta}_k \\ \overline{\eta}^2 \Delta_k & \overline{\eta}\eta(-i\omega_n - \chi_k - \lambda) \end{pmatrix} \right] \begin{pmatrix} c_{k\uparrow} \\ \overline{c}_{-k\downarrow} \end{pmatrix}. \quad (6)$$

The quasiparticle excitation energy spectrum $E_k$ is determined from the following equation: $E_k^2 \left[ E_k^2 - (\chi_k + \lambda)^2 - |\Delta_k|^2 + \overline{\eta}\eta \right]^2 = \{[E_k^2 - (\chi_k + \lambda)^2 - |\Delta_k|^2]\xi_k + \overline{\eta}\eta(\chi_k + \lambda)\}^2 + (\overline{\eta}\eta)^2 \overline{\Delta}_k \Delta_k$. This equation has the simple solution at the $J_K \to \infty$ limit, providing that $s_0 \neq 0$, as follows,

$$E_k = \pm\sqrt{(\chi_k + \lambda)^2 + \overline{\Delta}_k \Delta_k}. \quad (7)$$

This is the quasiparticle spectrum of $d_{x^2-y^2}$ wave superconductivity and corresponds to that of the slave boson mean field theory. On the other hand, the phase with $\eta = 0$ corresponds to the pseudogap phase of the slave boson mean field theory. In this phase, we find that $E_k = \xi_k$. So there is no excitation gap for the carriers. But there is the excitation energy gap with d-wave symmetry in the localized spin moment system. Therefore, we have obtained the same picture for the d-wave superconductive phase and the pseudogap phase.

## 3. Alternative formalism for the slave fermion mean field theory

Now we shall discuss an alternative formalism for the slave *fermion* mean field theory of the t-J model. The most crucial difference is that the absence of the $d_{x^2-y^2}$-wave superconductive state at the mean field theory as we shall see below. Instead of the Abrikosov pseudofermion, we use the Schwinger boson to represent the localized spin:

$$\mathbf{S}_j = \frac{1}{2} z_j^\dagger \boldsymbol{\sigma} z_j, \tag{8}$$

with $\langle \sum_{\sigma=\uparrow,\downarrow} z_{j\sigma}^\dagger z_{j\sigma} \rangle = 2\tilde{S}$. The Kondo coupling term is rewritten as $(\bar{c}_j \boldsymbol{\sigma} c_j) \cdot (\bar{z}_j \boldsymbol{\sigma} z_j) \to \frac{3}{2}\bar{\chi}_j \chi_j - \frac{3}{\sqrt{2}}\bar{\chi}_j (z_{j\downarrow} c_{j\uparrow} - z_{j\uparrow} c_{j\downarrow}) - \frac{3}{\sqrt{2}}\chi_j (\bar{c}_{j\uparrow} \bar{z}_{j\downarrow} - \bar{c}_{j\downarrow} \bar{z}_{j\uparrow})$. Here the SH fields for the triplet pair formation have been omitted as in the Abrikosov pseudofermion case. The SH fields $\chi_j$ and $\bar{\chi}_j$ are Grassmann numbers. The term of the Heisenberg model is rewritten as follows, $(\bar{z}_i \boldsymbol{\sigma} z_i) \cdot (\bar{z}_j \boldsymbol{\sigma} z_j) = -2(\bar{z}_{j\downarrow} \bar{z}_{i\uparrow} - \bar{z}_{j\uparrow} \bar{z}_{i\downarrow})(z_{i\uparrow} z_{j\downarrow} - z_{i\downarrow} z_{j\uparrow}) + \text{const}$. The action is given by

$$\begin{aligned}
S = \int_0^\beta d\tau \Bigg[ &\sum_j \bar{c}_j (\partial_\tau - \mu) c_j + \sum_j \bar{z}_j \partial_\tau z_j - t \sum_{\langle i,j \rangle} (\bar{c}_i c_j + \bar{c}_j c_i) + \sum_j \lambda_j (\bar{z}_j z_j - 2\tilde{S}) \\
&+ \frac{3}{4} J_K \sum_j \bar{\chi}_j \chi_j + \frac{J}{2} \sum_{\langle i,j \rangle} \bar{A}_{ij} A_{ij} \\
&- \frac{3}{4} J_K \sum_j \left[ \bar{\chi}_j (z_{j\downarrow} c_{j\uparrow} - z_{j\uparrow} c_{j\downarrow}) + \chi_j (\bar{c}_{j\uparrow} \bar{z}_{j\downarrow} - \bar{c}_{j\downarrow} \bar{z}_{j\uparrow}) \right] \\
&- \frac{J}{2} \sum_{\langle i,j \rangle} \left[ \bar{A}_{ij} (z_{i\uparrow} z_{j\downarrow} - z_{i\downarrow} z_{j\uparrow}) + A_{ij} (\bar{z}_{j\downarrow} \bar{z}_{i\uparrow} - \bar{z}_{j\uparrow} \bar{z}_{i\downarrow}) \right] \Bigg],
\end{aligned} \tag{9}$$

where SH fields $A_{ij}$ and $\bar{A}_{ij}$ are complex numbers.

We consider the uniform solution of $\bar{\chi}_j$, $\chi_j$, $\bar{A}_{ij}$, and $A_{ij}$ of the saddle point equations. This corresponds to the slave fermion mean field theory of the t-J model. In the momentum and the Matsubara frequency space, we obtain

$$\begin{aligned}
S = &\sum_k \bar{c}_k (-i\omega_n + \xi_\mathbf{k}) c_k + \sum_k \bar{z}_k (-i\omega_n + \lambda) z_k - \frac{3}{4} J_K \sum_k [\bar{\chi}_0 (z_{-k\downarrow} c_{k\uparrow} - z_{k\uparrow} c_{-k\downarrow}) \\
&+ \chi_0 (\bar{c}_{k\uparrow} \bar{z}_{-k\downarrow} - \bar{c}_{-k\downarrow} \bar{z}_{k\uparrow})] - A \sum_k \gamma_\mathbf{k} (z_{-k\downarrow} z_{k\uparrow} + \bar{z}_{k\uparrow} \bar{z}_{-k\downarrow}),
\end{aligned} \tag{10}$$

with $\gamma_{\mathbf{k}} = (\sin k_x \pm \sin k_y)/2$ where the plus sign is for $A_{i,i+\hat{x}} = A_{i,i+\hat{y}} = A$ and the minus sign is for $A_{i,i+\hat{x}} = -A_{i,i+\hat{y}} = A$. Integrating out Schwinger bosons yields,

$$S_{\text{eff}}^b = \frac{9}{16} J_K^2 \sum_k \frac{1}{(i\omega_n)^2 - \lambda^2 - A^2 \gamma_k^2} \begin{pmatrix} \bar{c}_{k\uparrow} & c_{-k\downarrow} \end{pmatrix}$$
$$\times \begin{pmatrix} \bar{\chi}_0 \chi_0 (i\omega_n - \lambda) & 0 \\ 0 & \bar{\chi}_0 \chi_0 (i\omega_n + \lambda) \end{pmatrix} \begin{pmatrix} c_{k\uparrow} \\ \bar{c}_{-k\downarrow} \end{pmatrix}. \tag{11}$$

It has been used that the square of a Grassmann variable, such as $\chi_0^2$ is zero. Note that there is no off-diagonal term at the mean field level. Therefore, the superconductive state is absent at least within the mean field theory.

## 4. Holon operator

In the above formalism, we have represented the localized spin either by the Abrikosov pseudofermion or by the Schwinger boson. This corresponds to represent the spinon by a fermion or a boson, respectively. Now we show that the SH fields $s_j$ and $\chi_j$ can be seen as the holon annihilation operator.

We first consider the case of the Abrikosov pseudofermion representation. The basis states of the Hilbert space at site $j$ are represented as $|c, f\rangle = |c\rangle \otimes |f\rangle$, where $|c\rangle$ denotes the carrier state and $|f\rangle$ denotes the localized spin state. We assume that the doubly occupied state at the copper site is excluded due to the strong on-site Coulomb repulsion. Then, the constraint is $\sum_\sigma f_{j\sigma}^\dagger f_{j\sigma} \leq 1$. Thus the basis of the Hilbert space is $|0, \uparrow\rangle$, $|0, \downarrow\rangle$, $|\downarrow, \uparrow\rangle$, $|\uparrow, \downarrow\rangle$, $|\uparrow, 0\rangle$, $|\downarrow, 0\rangle$, and $|0, 0\rangle$. Here 0 represents the unoccupied state, etc. In this basis, the state like $|\uparrow, \uparrow\rangle$ has been excluded because of the strong antiferromagnetic Kondo coupling.

By constructing the matrix representation of $s_j$, which has the form of $s_j = \frac{1}{\sqrt{2}}(f_{j\downarrow} c_{j\uparrow} - f_{j\uparrow} c_{j\downarrow})$ at the saddle point, in the above Hilbert space from those of $f_{j\sigma}$ and $c_{j\sigma}$, we find that the operation of $s_j$ does not change the total spin at site $j$ but annihilate the charge by one at site $j$. Therefore, $s_j$ is like the holon annihilation operator with bosonic statistics. However, the identification of $s_j$ with the holon operator is incomplete. For the full correspondence, we need to assume that the annihilation of the spin $\sigma$ spinon is equivalent to the creation of the spin $-\sigma$ spinon. This point may be more clarified if we formulate the theory in terms of the SU(2) doublets introduced in Ref.[24]. Representing the spinons in terms of those SU(2) doublets would give us an alternative formalism to the SU(2) formalism of the slave particle theory[25]. Similarly, we can show that $\chi_j$ is like the holon annihilation operator with fermionic statistics. The dynamics of the holons is determined by the action that is derived by integrating out the carrier fields and the localized spin moment fields. In case of the Abrikosov pseudofermion representation, the resulting action is $S_h = \sum_{\mathbf{q}, i\Omega_n} K(\mathbf{q}, i\Omega_n) \bar{\eta}(\mathbf{q}, i\Omega_n) \eta(\mathbf{q}, i\Omega_n)$, where $K(\mathbf{q}, i\Omega_n) = (1/\beta) \sum_{\mathbf{k}, i\omega_n} [i\omega_n(i\omega_n + i\Omega_n) - (\chi_{\mathbf{k}} + \lambda) \xi_{\mathbf{k+q}}] / \{[(i\omega_n)^2 - E_{\mathbf{k}}^2][(i\omega_n + i\Omega_n)^2 - \xi_{\mathbf{k+q}}^2]\}$.

## 5. Gauge field and deconfinement

Now we discuss the U(1) gauge field. The phase fluctuation of $\chi_{ij}$ corresponds to the U(1) gauge field considered in Refs.[7, 8]. Indeed, the gauge field is associated with the phase fluctuations of the same SH field for the spinons. As argued in Refs.[7, 8], the transition temperature, which is on the order of $J$ within the mean field analysis, could be suppressed by the gauge field fluctuations. Since the effective gauge field propagator depends on the dynamics of the holon, which is governed by $S_h$, some physical properties of the slave particle formalism, such as the $T$-linear resistivity law, could be different. This point is left for the future study.

Now we consider the deconfinement nature of the gauge field in the present formalism. Contrary to the slave boson theory[7, 8], the spinon fields and the carrier fields are independent fields because we have started from the spin-fermion like model in which the carriers and the localized spin moments are independent. In this sense the spinons are not confined to the electrons. However, one can show that in the Néel ordered phase the U(1) gauge field associated with the phase fluctuations of $\chi_{ij}$ leads to a logarithmic confining potential between the Abrikosov's pseudofermions. In other words, there is no low-lying spin 1/2 excitations in the localized spin system. Whereas in the disordered phase the logarithmic potential is replaced with a potential of exponential decay. Therefore, the gauge field is not confining in this phase and the fermions are no more confined to pairs.

## 6. Conclusion

In this paper, we have proposed an alternative formalism to the slave-particle theory of the t-J model. Here we have started from the spin-fermion model that is a multiband model and is reduced to the t-J model by taking the strong Kondo couping limit. The strong coupling limit is taken after introducing fields that describe the localized spin moments. Since the spin-fermion model is a multiband model, the carrier fields and the fields describing the localized spin moments are independent fields.

Main results concerning the superconductive states of the slave-particle theory at the mean field level are reproduced by the present formalism. The gauge field fluctuations lead to confinement phase of the spin-1/2 particles that describe the localized spin moments in the Néel ordering phase. Meanwhile, those particles are not confinend in the disordered phase.

### Acknowledgments

This work was partially supported by the Grant-in-Aid for the 21st Century COE "Center for Diversity and Universality in Physics" from the Ministry of Education, Culture, Sports, Science and Technology (MEXT) of Japan.